\documentclass[article,12pt,nofootinbib]{revtex4}
\usepackage[paper=letterpaper,margin=1.5in]{geometry}
\usepackage{epsfig,color}
\usepackage{graphicx}
\usepackage{amssymb,amsmath}
\usepackage{time}
\usepackage[varg]{txfonts}
\usepackage{hyperref}
\hypersetup{
  colorlinks,
  citecolor=green,
  linkcolor=blue}
\newcommand{\be}{\begin{equation}}
\newcommand{\ee}{\end{equation}}
\newcommand{\bc}{\begin{center}}
\newcommand{\ec}{\end{center}}

\renewcommand{\Re}{{\rm Re}}
\renewcommand{\Im}{{\rm Im}}

\renewcommand{\vec}[1]{\textnormal{\boldmath$#1$}}
\begin{document}

\bibliographystyle{revtex}


\vspace{.4cm}

\title
{Axial representation of external elctromagnetic fields for particle tracking in accelerators
}
\author{Igor Zagorodnov and Martin Dohlus}
\affiliation{Deutsches Elektronen-Synchrotron, Notkestrasse 85,
22603 Hamburg, Germany}
\date{\today}

\vspace{.4cm} 
\begin{abstract} 
We derive a power series representation of an arbitrary electromagnetic field near some  axis through the coaxial field components on the axis. The obtained equations are compared with Fourier-Bessel series approach and verified by several examples. It is shown that for each azimuthal mode  we need only two real functions on the axis in order to describe the field in a source free region near to it. The representation of dipole mode in a superconducting radio-frequency gun is analyzed.
\end{abstract}

\maketitle

\section{Introduction}
The modern particle accelerators based on radio-frequency technology use electromagnetic fields for the particle acceleration and the guiding~\cite{RF, Sec}. In order to study the beam dynamics numerically we need a representation of the external  fields near to the beam trajectory. 

In some ideal symmetric structures the fields can be described analytically. However, in most cases they are obtained numerically with help of electromagnetic solvers~\cite{CST} or measured experimentally. For accurate beam dynamics simulations we need a sampling of the fields on a three dimensional mesh with a high resolution.

In this paper we suggest an approach to describe an arbitrary time varying electromagnetic field as a superposition of several real functions on the reference trajectory only. Such approach is used already in code ASTRA~\cite{ASTRA}, for example. However, only monopoly (fully rotationally symmetric) harmonic fields are implemented with expansion up to the third order in radial coordinate. Our main interest is to generalize this technique on arbitrary electromagnetic fields  in  a source free region with a goal to use this compact and computationally efficient representation in beam dynamics codes~\cite{OCELOT,Krack3}.

The field power series representations are used in modern map-based beam dynamics codes to model magnetic elements~\cite{Dragt}. However, as noted in~\cite{Abell}, for the radio-frequency (rf) cavities the transfer maps are usually computed as either energy kicks or linear maps. In~\cite{Abell} a method for including the non-linear effects of rf cavities in a map-based code is introduced. However, the obtained equations are not local: they include Fourier transforms in time and space coordinates. The axial field representation is described in~\cite{Abell} only for monopole  field and a numerical analysis is done only for monopole field as well. 

The power series representation developed in this paper  is local:  it does not require any integral transformations or special functions.  For each azimuthal mode  we need  to know only two real functions on the axis in order to describe the field in a source free region near to it. 

We start in Section~\ref{SecEFE} with the Maxwell's equations in cylindrical coordinates and describe some helpful relations between the field components. In Section~\ref{SecFB} we consider an approach based on Fourier-Bessel series.  A simple argumentation and an example of analytical solenoid field in Section~\ref{Sec_Solenoid}  highlight an inaccuracy and  inefficiency of such approach in attempt to present the  electromagnetic fields through the coaxial field components on the axis. In Section~\ref{SecPS} we derive  another representation of electromagnetic fields based on power series expansion in radial cordinate. Section~\ref{SecEx}  contains three examples. The first example is the solenoid magnetic field. We use it to demonstrate that our equations reproduce the well known result published, for example, in \cite{Kareh}, and that the power series approach is more accurate with less computational efforts as compared to the method based on  Fourier-Bessel series. The second and the third examples are dipole harmonic fields.  We derive all equations in an explicit form and use them to present the dipole modes in a spherical cavity and in a superconducting rf gun~\cite{Sec}. The accuracy of the numerical resuts is analysed. The power series representation of the dipole mode in the gun is compared with the direct results from an eigenmode solver.

The equations are written in Gaussian units. 

\section{Electromagnetic field equations in cylindrical coordinates}\label{SecEFE}

In  cylindrical coordinates $(r,\theta,z)$ an arbitrary electromagnetic field $(\vec{E},\vec{B})$ near the $z$-axis can be represented as Fourier series azimuthal expansion
\begin{align}
\vec{E}(r,\theta,z,t)&=\Re \bigg(\sum_{m=0}^{\infty}\vec{E}^{(m)}(r,z,t) e^{i m \theta}\bigg),\quad \vec{E}^{(m)}=(E_{r}^{(m)},E_{\theta}^{(m)},E_{z}^{(m)})^T\nonumber,\\
\vec{B}(r,\theta,z,t)&=\Re \bigg(\sum_{m=0}^{\infty}\vec{B}^{(m)}(r,z,t) e^{i m \theta}\bigg),\quad \vec{B^{(m)}}=(B_{r}^{(m)},B_{\theta}^{(m)},B_{z}^{(m)})^T\nonumber.
\end{align}
Let us consider a source free domain around $z$-axis. The Maxwell's equations for the azimuthal harmonics at each azimuthal number $m$ read
\begin{align}\label{Maxwell1}
\frac{1}{r} \frac{\partial}{\partial r}(r B_r^{(m)})+\frac{im}{r}B_{\theta}^{(m)}+\frac{\partial}{\partial z}B_z^{(m)}=0,\\
\frac{1}{r} \frac{\partial}{\partial r}(r E_r^{(m)})+\frac{im}{r}E_{\theta}^{(m)}+\frac{\partial}{\partial z}E_z^{(m)}=0, \nonumber \\
\frac{im}{r}B_z^{(m)}-\frac{\partial}{\partial z}B_{\theta}^{(m)}=\frac{\partial}{c\partial t} E_r^{(m)}, \nonumber \\
\frac{\partial}{\partial z}B_r^{(m)}-\frac{\partial}{\partial r}B_z^{(m)}= \frac{\partial}{c\partial t}  E_{\theta}^{(m)}, \nonumber \\
\frac{1}{r} \frac{\partial}{\partial r}(r B_{\theta})-\frac{im}{r}B_{r}= \frac{\partial}{c\partial t}  E_z,\nonumber  \\
\frac{im}{r}E_z^{(m)}-\frac{\partial}{\partial z}E_{\theta}^{(m)}=-\frac{\partial}{c\partial t} B_r^{(m)}, \nonumber \\
\frac{\partial}{\partial z}E_r^{(m)}-\frac{\partial}{\partial r}E_z^{(m)}= - \frac{\partial}{c\partial t}  B_{\theta}^{(m)}, \nonumber \\
\frac{1}{r} \frac{\partial}{\partial r}(r E_{\theta}^{(m)})-\frac{im}{r}E_{r}^{(m)}= - \frac{\partial}{c\partial t}  B_z^{(m)},\nonumber 
\end{align}
where $c$ is the speed of light in vacuum. Taking the real and the imaginary parts of equations in system~(\ref{Maxwell1}) we separete them into two independent sets of equations. Each of the two systems can be written in the form 
\begin{align}\label{Maxwell2}
&\frac{1}{r} \frac{\partial}{\partial r}(r b_r)+\frac{M}{r}b_{\theta}+\frac{\partial}{\partial z}b_z=0,\quad
\frac{1}{r} \frac{\partial}{\partial r}(r e_r)-\frac{M}{r}e_{\theta}+\frac{\partial}{\partial z}e_z=0,  \\
&-\frac{M}{r}b_z-\frac{\partial}{\partial z}b_{\theta}=\frac{\partial}{c\partial t} e_r, \quad
\frac{\partial}{\partial z}b_r-\frac{\partial}{\partial r}b_z= \frac{\partial}{c\partial t}  e_{\theta}, \quad
\frac{1}{r} \frac{\partial}{\partial r}(r b_{\theta})+\frac{M}{r}b_{r}= \frac{\partial}{c\partial t}  e_z,\nonumber  \\
&\frac{M}{r}e_z-\frac{\partial}{\partial z}e_{\theta}=-\frac{\partial}{c\partial t} b_r, \quad
\frac{\partial}{\partial z}e_r-\frac{\partial}{\partial r}e_z= - \frac{\partial}{c\partial t}  b_{\theta}, \quad
\frac{1}{r} \frac{\partial}{\partial r}(r e_{\theta})-\frac{M}{r}e_{r}= - \frac{\partial}{c\partial t}  b_z,\nonumber 
\end{align}
where $M=m$ for for the first set of variables 
\begin{align}\
e_r=\Re( E_r^{(m)}),\quad e_{\theta}=\Im( E_{\theta}^{(m)}),\quad e_z=\Re(E_z^{(m)}),\nonumber\\
b_r=\Im(B_r^{(m)}),\quad b_{\theta}=\Re(B_{\theta}^{(m)}),\quad b_z=\Im(B_z^{(m)}),\nonumber 
\end{align}
and $M=-m$ for  the second set of variables
\begin{align}\
e_r=\Im( E_r^{(m)}),\quad e_{\theta}=\Re( E_{\theta}^{(m)}),\quad e_z=\Im(E_z^{(m)}),\nonumber\\
b_r=\Re(B_r^{(m)}),\quad b_{\theta}=\Im(B_{\theta}^{(m)}),\quad b_z=\Re(B_z^{(m)}).\nonumber 
\end{align}
We will call the first set of variables "$M^+$" field and the second set   will be denoted as "$M^-$" field. 

From Eq.~(\ref{Maxwell2}) we can derive the second order wave equations for the longitudinal components only
\begin{align}\label{WaveLong}
\frac{1}{r} \frac{\partial}{\partial r}(r \frac{\partial}{\partial r} e_z)-\frac{m^2}{r^2}e_z=\Box  e_z, \quad
\frac{1}{r} \frac{\partial}{\partial r}(r \frac{\partial}{\partial r} b_z)-\frac{m^2}{r^2}b_z=\Box b_z, 
\end{align}
where symbol $\Box$  denotes the  d'Alambert operator~\cite{Vladimirov}
\begin{align}\
\Box\equiv\frac{\partial^2}{c^2\partial t^2}- \frac{\partial^2}{\partial z^2}.\nonumber 
\end{align}
Each of the transverse field components can be related to the longitudinal field components  through the relations 
\begin{align}\label{WaveTrans}
e_r=-\Box^{-1}\bigg[\frac{M}{r}\frac{\partial}{c\partial t} b_z+\frac{\partial^2}{\partial r \partial z} e_z\bigg],\quad
e_{\theta}=-\Box^{-1}\bigg[\frac{\partial^2}{c\partial t\partial r} b_z+\frac{M}{r}\frac{\partial}{\partial z} e_z\bigg],\\ 
b_r=-\Box^{-1}\bigg[\frac{M}{r}\frac{\partial}{c\partial t} e_z+\frac{\partial^2}{\partial r \partial z} b_z\bigg],\quad
b_{\theta}=\Box^{-1}\bigg[\frac{\partial^2}{c\partial t\partial r} e_z+\frac{M}{r}\frac{\partial}{\partial z} b_z\bigg].\nonumber 
\end{align}
where $\Box^{-1}$  is an integral operator inverse to the d'Alambert operator $\Box$. This integral can be written with the help of Green's function~\cite{Vladimirov} when the initial and the boundary conditions are given. However, we do not need the explicit form of this operator as we only will use the property $\Box^{-1}\Box=1$.  

\section{Axial representation of elctromagnetic fields through Fourier-Bessel series}\label{SecFB}

In this section we will assume that the fields have support only in interval $-Z\le z<Z$. 
A general solution of the wave equations~(\ref{WaveLong}) for the longitudinal field components can be written as Fourier-Bessel series
\begin{align}\label{EqFBZ}
e_z(r,z,t)&=\int_{-\infty}^{\infty}dk e^{i k c t}\sum_{p=-\infty}^{\infty} e^{i k_{z,p} z} e_{z,p}(k) J_m(\nu_{z,p}r),\\
b_z(r,z,t)&=\int_{-\infty}^{\infty}dk e^{i k c t}\sum_{p=-\infty}^{\infty} e^{i k_{z,p} z} b_{z,p}(k) J_m(\nu_{z,p}r),\nonumber\\
\quad k_{z,p}&=\pi\frac{p}{Z},\quad \nu_{z,p}=\sqrt{k^2-k_{z,p}^2}.
\nonumber
\end{align}
The transverse field components read
\begin{align}\label{EqFBT}
e_{\theta}(r,z,t)&=\int_{-\infty}^{\infty}dk e^{i k c t}\sum_{p=-\infty}^{\infty} e^{i k_{z,p} z} e_{\theta,p}(r,k),\\
e_r(r,z,t)&=\int_{-\infty}^{\infty}dk e^{i k c t}\sum_{p=-\infty}^{\infty} e^{i k_{z,p} z} e_{r,p}(r,k),\nonumber \\
b_{\theta}(r,z,t)&=\int_{-\infty}^{\infty}dk e^{i k c t}\sum_{p=-\infty}^{\infty} e^{i k_{z,p} z} b_{\theta,p}(r,k),\nonumber\\
b_r(r,z,t)&=\int_{-\infty}^{\infty}dk e^{i k c t}\sum_{p=-\infty}^{\infty} e^{i k_{z,p} z} b_{r,p}(r,k).
\nonumber
\end{align}
From Eq.~(\ref{WaveTrans}) we obtain the representation of the coefficients of the transverse fields through the coefficients of the longitudinal fields 
\begin{align}\label{EqFBC}
e_{r,p}(r,k)=\frac{i}{\nu_{z,p}^2}\bigg[ \frac{M}{r} k b_{z,p}(k)J_m(\nu_{z,p}r)+k_z e_{z,p}(k)\frac{\partial}{\partial r} J_m(\nu_{z,p}r)  \bigg],\\
e_{\theta,p}(r,k)=\frac{i}{\nu_{z,p}^2}\bigg[ k b_{z,p}(k)\frac{\partial}{\partial r} J_m(\nu_{z,p}r)+ \frac{M}{r} k_z e_{z,p}(k)J_m(\nu_{z,p}r)  \bigg],
\nonumber\\
b_{r,p}(r,k)=\frac{i}{\nu_{z,p}^2}\bigg[ \frac{M}{r} k e_{z,p}(k)J_m(\nu_{z,p}r)+k_z b_{z,p}(k)\frac{\partial}{\partial r} J_m(\nu_{z,p}r)  \bigg],
\nonumber\\
b_{\theta,p}(r,k)=-\frac{i}{\nu_{z,p}^2}\bigg[ k e_{z,p}(k)\frac{\partial}{\partial r} J_m(\nu_{z,p}r)+ \frac{M}{r} k_z b_{z,p}(k)J_m(\nu_{z,p}r)  \bigg].
\nonumber
\end{align}

In this paper we are interested in representation of the fields near some axis through the field values on the axis. In order to be able to present an arbitrary field components (or their derivatives) on the axis we have to take  $k_{z,p}$ to be real. Hence if the field has harmonics with $|k|<|k_{z,p}|$ then the Bessel functions will have an imaginary argument and will show an exponential growth for large arguments. In finite-digit  arithmetic of the computer such representation will be accurate only in small vicinity of the axis and will deverge fast with increasing of the order of harmonics or the distance from the axis. We will confirm it by an analytical example of solenoid field in Section~\ref{Sec_Solenoid}. Additional drawback of this representation is usage of large number of different special functions at each point of interest. Such kind of interpolaion requires considerable computational time.

In the next Section we will work out an another representation which allows an accurate description of electromagnetic fields up to boundaries of the source free domain with small number of arithmetic operations.

\section{Axial representation of elctromagnetic fields through power series}\label{SecPS}

We will seek the solution for all field components in the form of the Taylor series expansion near the axis. For example, the longitudinal components of the fields read
\begin{align}
\label{ExpLong}
e_z(r,z,t)=\sum_{n=0}^{\infty}e_{z,n}r^n,\quad b_z(r,z,t)=\sum_{n=0}^{\infty}b_{z,n}r^n.  
\end{align}
A similar form of the expansion is used for the transverse field components as well. If we put Eq.~(\ref{ExpLong}) into Eq.~(\ref{WaveLong}) we will find out that the expansion coefficients fulfill the recursive relations
\begin{align} 
\label{RecLong}
e_{z,n}= \frac{\Box e_{z,n-2}}{n^2-m^2},\quad b_{z,n}= \frac{\Box b_{z,n-2}}{n^2-m^2}. 
\end{align}
In the same manner we derive from  Eq.~(\ref{WaveTrans}) a relation relating the radial electric field expansion coefficients to the longitudinal ones
\begin{align}
e_{r,n}=-\Box^{-1}\bigg[M\frac{\partial}{c\partial t} b_{z,n+1}+(n+1)\frac{\partial}{\partial z} e_{z,n+1}\bigg].\nonumber
\end{align}
In order to remove the integral operator $\Box^{-1}$ from the last equation we use the recursive relation Eq.~(\ref{RecLong}) and obtain
\begin{align}\label{TransEr}
e_{r,n}=-\frac{1}{(n+1)^2-m^2}\bigg[M\frac{\partial}{c\partial t} b_{z,n-1}+(n+1)\frac{\partial}{\partial z} e_{z,n-1}\bigg].
\end{align}
Likewise the coefficients of other transverse components can be written as
\begin{align}\label{TransOther}
e_{\theta,n}&=-\frac{1}{(n+1)^2-m^2}\bigg[(n+1)\frac{\partial}{c\partial t} b_{z,n-1}+M\frac{\partial}{\partial z} e_{z,n-1}\bigg],\\ 
b_{r,n}&=-\frac{1}{(n+1)^2-m^2}\bigg[M\frac{\partial}{c\partial t} e_{z,n-1}+(n+1)\frac{\partial}{\partial z} b_{z,n-1}\bigg],\nonumber\\ 
b_{\theta,n}&=\frac{1}{(n+1)^2-m^2}\bigg[(n+1)\frac{\partial}{c\partial t} e_{z,n-1}+M\frac{\partial}{\partial z} b_{z,n-1}\bigg].\nonumber
\end{align}

Let us consider a field with azimuthal number  $m> 0$. From Eq.~(\ref{Maxwell2}) we can derive the initial conditions on the expansion coefficients
\begin{align}\label{Init}
e_{z,n}&=0,\quad    b_{z,n}=0,\quad n=1,2,..,m-1,  \\ 
e_{z,m}&=\bigg[\frac{1}{m}\frac{\partial}{\partial z} e_{r,m-1}-\frac{1}{M}\frac{\partial}{c\partial t}b_{r,m-1}\bigg],\quad
b_{z,m}=\bigg[\frac{1}{m}\frac{\partial}{\partial z} b_{r,m-1}-\frac{1}{M}\frac{\partial}{c\partial t}e_{r,m-1}\bigg], \nonumber\\
e_{r,m-1}&=\frac{1}{(m-1)!}\frac{\partial^{m-1}}{\partial r^{m-1}} e_{r}(0,z,t),\quad b_{r,m-1}=\frac{1}{(m-1)!}\frac{\partial^{m-1}}{\partial r^{m-1}} b_{r}(0,z,t), \nonumber\\
e_{\theta,m-1}&=\frac{M}{m}e_{r,m-1},\quad b_{\theta,m-1}=-\frac{M}{m}b_{r,m-1}. \nonumber
\end{align}
For  $m$ even the longitudinal field components have the expansion only in even powers of $r$:
\begin{align}\label{EqZeven}
e_z(r,z,t)=\sum_{n=m/2}^{\infty} e_{z,2n}(z,t)r^{2n},\quad b_z(r,z,t)=\sum_{n=m/2}^{\infty} b_{z,2n}(z,t)r^{2n},
\end{align}
where the coefficients are related by recursive Eq.~(\ref{RecLong}). The transverse field components can be written as expansion only in non-even powers of $r$:
\begin{align}\label{EqTeven}
e_r(r,z,t)=\sum_{n=m/2+1}^{\infty} e_{r,2n-1}(z,t)r^{2n-1},\quad e_{\theta}(r,z,t)=\sum_{n=m/2+1}^{\infty} e_{\theta,2n-1}(z,t)r^{2n-1},\\
b_r(r,z,t)=\sum_{n=m/2+1}^{\infty} b_{r,2n-1}(z,t)r^{2n-1},\quad b_{\theta}(r,z,t)=\sum_{n=m/2+1}^{\infty} b_{\theta,2n-1}(z,t)r^{2n-1},\nonumber 
\end{align}
with coefficients given by Eq.~(\ref{TransEr}),~(\ref{TransOther}). For the azimuthal number $m$ non-even the longitudinal components have an expansion only in non-even powers of $r$ and the transverse components of the electric filed can be expanded only in even powers of $r$. The coefficients can be found as in the former case from Eq.~(\ref{RecLong})-(\ref{TransOther}).

In order to describe "$M^+$" part of the field near the axis for azimuthal number  $m>0$ we need only 2 real functions  $e_{r,m-1}^{+}(0,z,t),  b_{r,m-1}^{+}(0,z,t)$ on the axis, which are related to the total field by relations  
  \begin{align}
e_{r,m-1}^{+}(0,z,t)=\int_{0}^{2\pi} \frac{\partial^{m-1}}{\partial r^{m-1}} E_{r}(0,\theta,z,t) \cos(m \theta)\frac{d\theta}{\pi},\nonumber\\
b_{r,m-1}^{+}(0,z,t)=-\int_{0}^{2\pi} \frac{\partial^{m-1}}{\partial r^{m-1}} B_{r}(0,\theta,z,t) \sin(m \theta)\frac{d\theta}{\pi}. \nonumber  
\end{align}
Likewise in order to describe "$M^-$" part of the field near the axis we need yet another 2 real functions  $e_{r,m-1}^{-}(0,z,t),  b_{r,m-1}^{-}(0,z,t)$  on the axis, which are related to the total field by similar relations 
\begin{align}
e_{r,m-1}^{-}(0,z,t)=-\int_{0}^{2\pi} \frac{\partial^{m-1}}{\partial r^{m-1}} E_{r}(0,\theta,z,t) \sin(m \theta)\frac{d\theta}{\pi},\nonumber \\
b_{r,m-1}^{-}(0,z,t)=\int_{0}^{2\pi} \frac{\partial^{m-1}}{\partial r^{m-1}} B_{r}(0,\theta,z,t) \cos(m \theta)\frac{d\theta}{\pi}. \nonumber
 \end{align}
Hence the harmonic of the full field for the azimuathal number $m>0$ can be written as direct sum of $M+$ and $M-$ fileds:
\begin{align}
\vec{E}=\vec{E}^{+}+\vec{E}^{-},\quad \vec{B}=\vec{B}^{+}+\vec{B}^{-},\nonumber
\end{align}
and the full description of azimuthal harmonic with $m>0$  near an arbitrary axis requires 4 real  functions on the axis.

The monopole fields, with azimuthal number $m=0$, have another form of conditions for the initial expansion coefficients:
\begin{align}\label{ICmono}
e_{r,0}=0,\quad b_{\theta,0}=0,\quad   e_{z,0}=e_z(0,r,t),\quad e_{z,1}=0, \\
b_{r,0}=0,\quad e_{\theta,0}=0,\quad   b_{z,0}=b_z(0,r,t),\quad b_{z,1}=0.  \nonumber
\end{align}
The higher order coefficients can be found in the same manner as before from Eq.~(\ref{RecLong})-(\ref{TransOther}).
In order to describe  the TM field with non-zero field components ($E_r,E_z, B_{\theta}$) we need only one real function on the axis: $E_{z}(0,z,t)$. Likewise the TE field can be described with only one function $B_{z}(0,z,t)$ on the axis.

The axial power series representation introduced above is local: it do not use any integral transformations in space or in time. For each arbitrary fixed point $(r_0,\theta_0,z_0,t_0)$  the field expansion coefficients depend  only on the field in infinitesimally small area near to the point $(0,\theta_0,z_0,t_0)$ on the axis. Hence Eq.~(\ref{RecLong})-(\ref{TransOther}) are correct for any kind of boundary or initial conditions in a source free region of interest.

The representation witn infinie sums is exact for any solution of the Maxwell's equations in a source free region. However, we are working with finite sums in the finite-digit computer arithmetic. Additionally, the power series representation obtained in this section requires to calculate the higher order derivatives of functions on the axis. In the finite-digit computer arithmetic or for data with errors it will cause a large inaccuracy and high frequency oscillations of large amplitudes. We will analyze these issues in the next section on several examples calculated with 64 bit floating point arithmetic. 

\section{Examples of axal representation of fields and error analysis}\label{SecEx}
\subsection{Field of finite solenoid}\label{Sec_Solenoid}

As our first example we consider the magnetic field of solenoid. The field is static and axially symmetric.  We  write below the explicit form of expansions in Fourier-Bessel series of Section~\ref{SecFB} and in power series of Section~\ref{SecPS}. The both expansions will be analyzed and compared on the analytical example of thin finite solenoid.

\begin{figure}[htbp]
	\centering
	\includegraphics*[height=70mm]{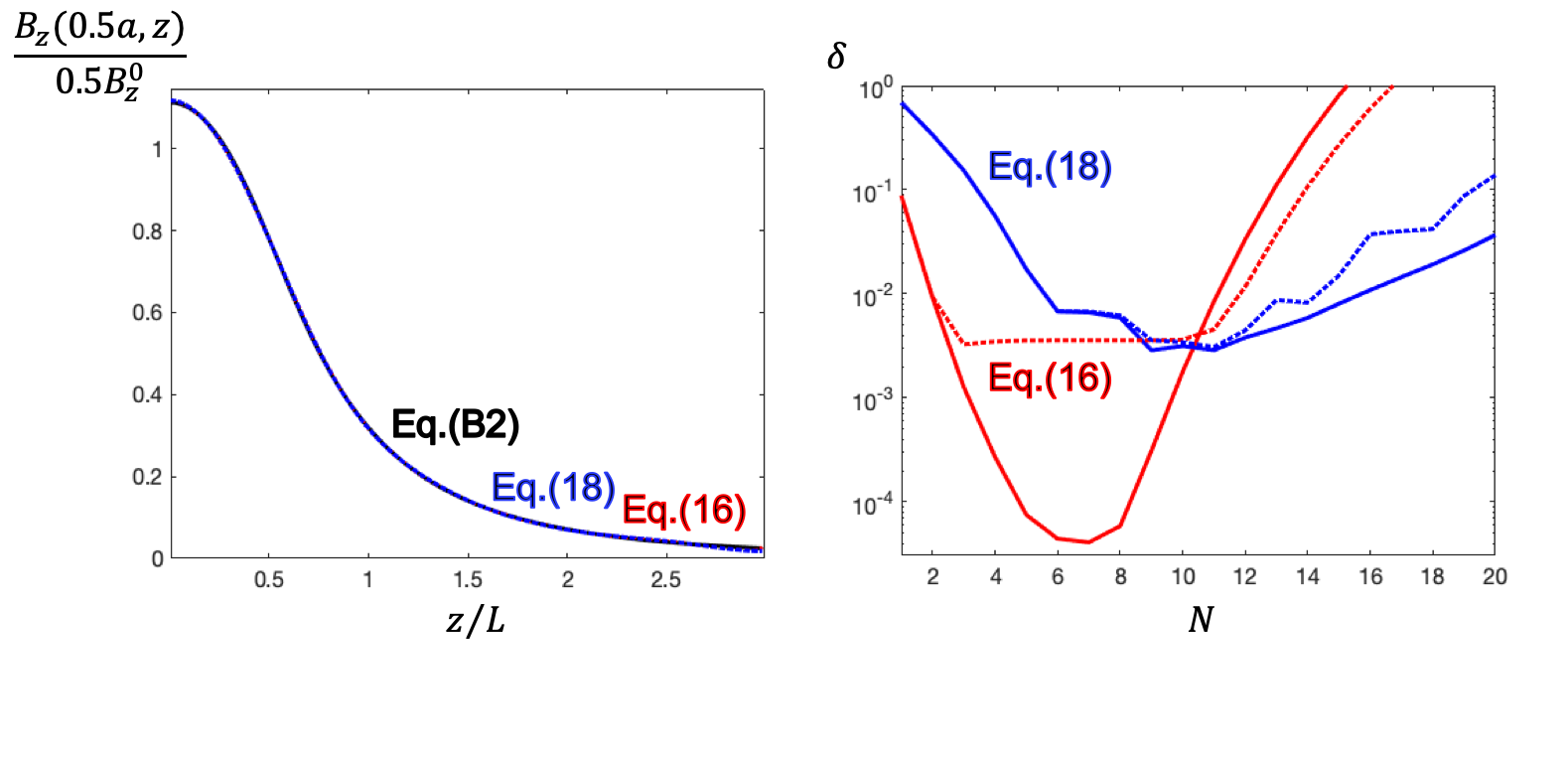}
	\caption{The left plot presents the longitudinal field component at the half radius of the solenoid. The right plot shows the error of Fourier-Bessel series (in blue) and of power series (in red) versus parameter $N$ in the series. The dashed curves present the same results when the analytical data on the axis are spoiled by a Gaussian distribution with rms of 1\%.}\label{Fig001}
\end{figure}

\begin{figure}[htbp]
	\centering
	\includegraphics*[height=70mm]{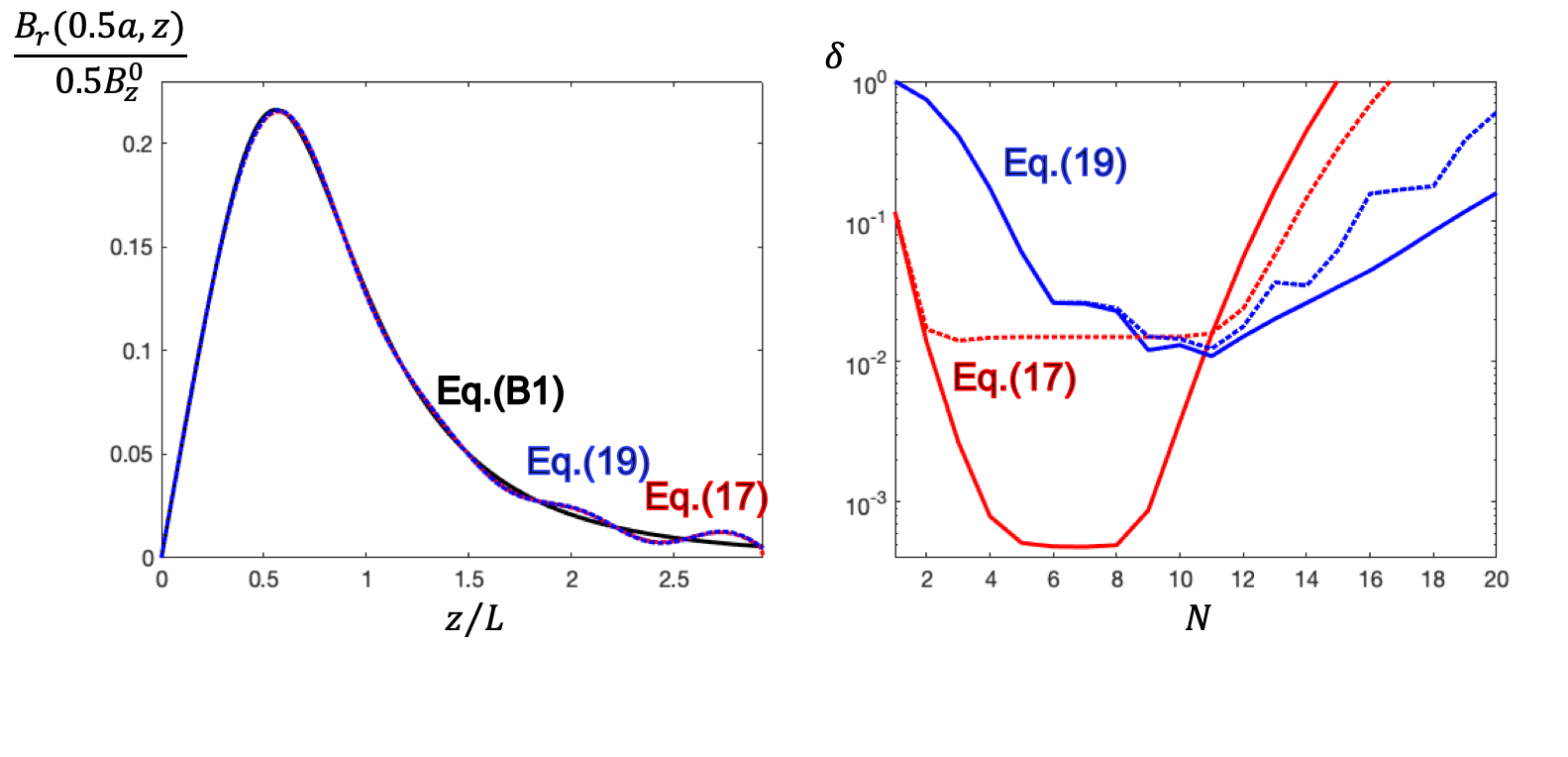}
	\caption{The left plot presents the radial field component at the half radius of the solenoid. The right plot shows the error of Fourier-Bessel series (in blue) and of power series (in red) versus parameter $N$ in the series. The dashed curves present the same results when the analytical data on the axis are spoiled by a Gaussian distribution with rms of 1\%.}\label{Fig002}
\end{figure}

The power series expansion can be obtained from Eq.~(\ref{EqZeven})-(\ref{ICmono}) for the monopole TE mode if we take $\frac{\partial}{c\partial t}=0$. The longitudinal magnetic field can be expanded as
\begin{align}\label{EqBzPS} 
B_z^{(N)}(r,z)=\sum_{n=0}^{N-1} b_{z,2n}(z)r^{2n}
\end{align}
with the recursive relation for the expansion coefficients
\begin{align}
b_{z,2n}=- \frac{1}{4n^2}\frac{\partial^2}{\partial z^2}b_{z,2n-2}.\nonumber 
\end{align}
From the latter relation the  representation of the coefficients through the field on the axis reads
\begin{align}
b_{z,0}=b_z(0,z),\quad
b_{z,2}=- \frac{1}{4}\frac{\partial^2}{\partial z^2}b_z(0,z),\quad
b_{z,2n}= \frac{(-1)^n}{(n!)^2 4^n} \frac{\partial^{2n}}{\partial z^{2n}} b_z(0,z).\nonumber 
\end{align}
The radial  magnetic field  component expands in non-even powers of $r$
\begin{align}\label{EqBrPS} 
B_r^{(N)}(r,z)=\sum_{n=1}^{N} b_{r,2n-1}(z)r^{2n-1},
\end{align}
with coefficients
\begin{align}
b_{r,1}=- \frac{1}{2}\frac{\partial}{\partial z}b_z(0,z),\quad
b_{r,3}=- \frac{1}{4}\frac{\partial}{\partial z}b_{z,2},\quad
b_{r,2n-1}=-\frac{1}{2n}\frac{\partial}{\partial z} b_{z,2n-2}.\nonumber
\end{align}
The representations obtained coincide with the ones published earlier, for example, in~\cite{Kareh}.

Let us now to consider the expansion through the Fourier-Bessel series. For the monopole field Eq.~(\ref{EqFBZ}),  (\ref{EqFBT})  reduce to the form 
\begin{align}\label{EqBzB}
B_z^{(N)}(r,z)&=\sum_{p=-N}^{N} e^{i k_{z,p} z} b_{z,p} J_0(\nu_{z,p}r),\\
\label{EqBrB}
B_r^{(N)}(r,z)&=\sum_{p=-N}^{N} e^{i k_{z,p} z} b_{r,p}(r),
\quad \nu_{z,p}^2=-k_{z,p}^2.
\end{align}
The expansion coefficients can be found through the relations 
\begin{align}
b_{z,p}= \frac{1}{2Z}\int_{-Z}^{Z}B_z(0,z)e^{-ik_{z,p}z}dz,\quad
b_{r,p}(r)= b_{z,p}J_1(\nu_{z,p}r),\nonumber
\end{align}
where the last equation is obtained from Eq.~(\ref{EqFBC}) with $m=0, k=0$.

If the field $B_z$ is known in some transverse plane (for example at plane with $z=0$ ) then it is possible to take $\nu_{z,p}$ as real numbers. However, we are interested in a representation through the field known only on the axis. Hence we have to choose $k_{z,p}=\pi\frac{p}{Z}$  as real numbers and, as consequence, the numbers  $\nu_{z,p}$ are imaginary ones. Hence the Bessel functions will show an exponential growth for large arguments.

We apply the both representations to the field of thin solenoid of length $L$. The analytical form of the solution was derived in~\cite{Callaghan, Haas} and can be found in Appendix B. The solenoid has the radius $a=\frac{\sqrt{3}}{2}L$  and the maximal longitudinal field on the axis is equal to $0.5 B_z^0$, where  $B_z^0$ is the field of an infinite solenoid.

The left plot in Fig.~\ref{Fig001} presents the longitudinal field component $B_z$ at the half radius of the solenoid. The right plot  shows the relative error 
\begin{align}
\delta=\frac{||B_z-B_z^{(N)}||_2}{||B_z||_2}\nonumber
\end{align}
of Fourier-Bessel series, Eq.~(\ref{EqBzB}), (in blue) and of power series, Eq.~(\ref{EqBzPS}),  (in red) versus the parameter $N$ in the series. The solid curves show the results  when the accurate analytical solution, Eq.~(\ref{ABzSol}), is sampled on the axis in interval $-3 L<z<3 L$ with step $0.02 L$.  The power series representation, Eq.~(\ref{EqBzPS}),  (red solid curve) reaches a high accuracy but after $N=7$ terms the accuracy starts to degrade as the higher order numerical derivatives are not accurate due to finite-digit computer arithmetic. The error of Fourier-Bessel series, Eq.~(\ref{EqBzB}), (solid blue curve) reaches the minimum with parameter $N=9$ (19 terms in the sum) and after this begins to degrade too due to an inaccuracy of the numerical values of the Bessel functions of large imaginary arguments.

In order to study an impact of errors due to measurements or calculations we have spoiled the accurate data by a random Gaussian distribution with rms deviation of 1 \%. The error obtained with Fourier-Bessel series, Eq.~(\ref{EqBzB}), for the spoiled data is shown by blue dashed curve. It reaches the same accuracy with parameter $N=9$ as in the former case with accurate sampling data.  However, the power series, Eq.~(\ref{EqBzPS}), cannot be applied to the spoiled data directly. The numerical approximations of derivatives result in large inaccuracies. Hence we need to remove the high frequency components in the spoiled data. It can be done, for example, with low-pass linear filter~\cite{DSP}. We used another approach: the function on the axis was Fourier transformed and recreated  with the same number of low order Fourier harmonics as used in  Fourier-Bessel series, Eq.~(\ref{EqBzB}),  with parameter $N=9$ (19 Fourier harmonics). The result is shown by the red dashed curve in the right plot. 

The left plot in Fig.~\ref{Fig001} shows three curves: the analytical  longitudinal field, Eq.~(\ref{ABzSol}), ( black solid line); the field calculated by Fouruer-Bessel series with $N=9$ from the spoiled data (blue dashed line);  the field calculated by power series with $N=3$ from the spoiled filtered data (red dashed line). The curves agree well with only small deviations at the ends of the interval.

Fig.~\ref{Fig002} presents the results for the radial field $B_r$ at the half radius of the solenoid. The curves and the colours have the same meaning as in Fig.~\ref{Fig001} for the longitudinal field. Again we see in the left plot very good agreement of the two methods with a small deviations at the ends of the interval.

It  can be concluded that the power series approach requires only $N=3$ terms to reach the same accuracy level as obtained by Fourier-Bessel series approach. The number of arithmetic operations  for the power series approach is negligible as compared to that required for the calcuation of many special functions in the Fourier-Bessel series approach.

\subsection{A harmonic field near the axis}

In this section we consider the power series representation of Section~\ref{SecPS} for electromagnetic fileds with harmonic dependence on time:
\begin{align}\label{EqHF}
\vec{E}(r,\theta,z,t)=\vec{E}(r,\theta,z)\cos(k c t),\quad \vec{B}(r,\theta,z,t)=\vec{B}(r,\theta,z)\sin(k c t).
\end{align}
For the harmonic fileds with such time dependence the equations of  Section~\ref{SecPS} should be modified in the following way: we have to replace the time derivative operator $\frac{\partial}{c\partial t}$ by $-k$ if it stands before an electric field component or by  $k$ if it stands before a magnetic field component.

Let us write the equations for the dipole mode, $m=1$. The relations below will describe the two decoupled modes simultaneously: the first mode can be obtained with $M=1$ and the second mode corresponds to $M=-1$.  

\begin{figure}[htbp]
	\centering
	\includegraphics*[height=50mm]{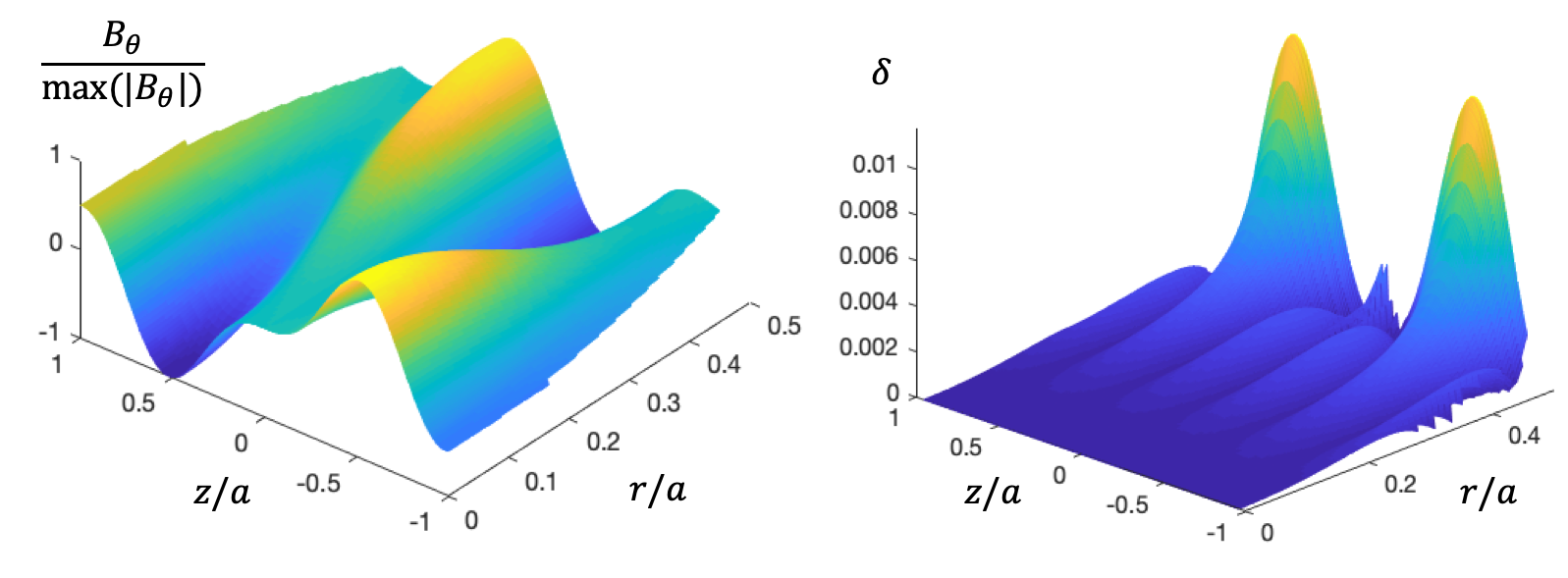}
	\caption{The left plot shows the azimuthal magnetic field component in the perfectly conducting sphere as obtained by Eq.~(\ref{DipTransB}) with $N=6$ coefficients. The plot  on the right shows the error, Eq.~(\ref{de}).}\label{Fig003}
\end{figure}

The Taylor series expansion of the longitudinal field components near the axis $z$ has only non-even powers
\begin{align}\label{DipLong}
E_z^{(N)}(r,z)=\sum_{n=0}^{N-1} e_{2n+1}(z)r^{2n+1},\quad B_z^{(N)}(r,z)=\sum_{n=1}^{N-1} b_{2n+1}(z)r^{2n+1}.
\end{align}
with the expansion coefficients defined by relations
\begin{align}
 e_{r,0}&=E_r(0,z),\quad b_{r,0}=B_r(0,z),\quad
 e_{\theta,0}=Me_{r,0},\quad b_{\theta,0}=-Mb_{r,0},\nonumber\\
e_{z,1}&=\bigg[\frac{\partial}{\partial z} e_{r,0}-M k b_{r,0}\bigg],\quad
b_{z,1}=\bigg[\frac{\partial}{\partial z} b_{r,0}+M k e_{r,0}\bigg],\nonumber\\
e_{z,2n+1}&= \frac{\bigg[-k^2- \frac{\partial^2}{\partial z^2}\bigg]^n e_{z,1}}{\prod_{p=1}^n [(2p+1)^2-1]},\quad
b_{z,2n+1}= \frac{\bigg[-k^2- \frac{\partial^2}{\partial z^2}\bigg]^n b_{z,1}}{\prod_{p=1}^n [(2p+1)^2-1]},\nonumber 
\end{align}
where we use Eq.~(\ref{Init}) and assume that the radial field components on the axis are known. The power series representations of the transverse field components in the whole source free domain have only even powers in radial coordinate
\begin{align}\label{DipTransE}
E_r^{(N)}(r,z)(r,z)=\sum_{n=0}^{N-1} e_{r,2n}(z,t)r^{2n},\quad E_{\theta}^{(N)}(r,z)=\sum_{n=0}^{N-1} e_{\theta,n}(z,t)r^{2n},\\
\label{DipTransB}
B_r^{(N)}(r,z)(r,z)=\sum_{n=0}^{N-1} b_{r,2n}(z,t)r^{2n},\quad B_{\theta}^{(N)}(r,z)=\sum_{n=0}^{N-1} b_{\theta,n}(z,t)r^{2n},
\end{align}
with the coefficients expressed through the expansion coefficients of the longitudinal field components only
\begin{align}\label{trans}
e_{r,2n}=-\frac{1}{(2n+1)^2-1}\bigg[M k b_{z,n-1}+(2n+1)\frac{\partial}{\partial z} e_{z,n-1}\bigg],\\ 
e_{\theta,2n}=-\frac{1}{(2n+1)^2-1}\bigg[(2n+1)k b_{z,n-1}+M\frac{\partial}{\partial z} e_{z,n-1}\bigg],\nonumber\\ 
b_{r,2n}=-\frac{1}{(2n+1)^2-1}\bigg[-M k e_{z,n-1}+(2n+1)\frac{\partial}{\partial z} b_{z,n-1}\bigg],\nonumber\\ 
b_{\theta,2n}=\frac{1}{(2n+1)^2-1}\bigg[-(2n+1)k e_{z,n-1}+M \frac{\partial}{\partial z} b_{z,n-1}\bigg].\nonumber
\end{align}
The coefficients of the field representation in explicit form up to fourth order read
\begin{align}
e_{z,1}&=\bigg[ \frac{\partial}{\partial z} e_{r,0}- M kb_{r,0}\bigg],\quad
e_{z,3}= \frac{\bigg[-k^2- \frac{\partial^2}{\partial z^2}\bigg] e_{z,1}}{8},\nonumber\\
b_{z,1}&=\bigg[ \frac{\partial}{\partial z} b_{r,0}+ M ke_{r,0}\bigg],\quad
b_{z,3}= \frac{\bigg[-k^2- \frac{\partial^2}{\partial z^2}\bigg] b_{z,1}}{8},\nonumber\\
 e_{\theta,0}&=M e_{r,0},\nonumber\\
e_{\theta,2 }&=- \frac{ \bigg[3 k b_{z,1}+M\frac{\partial}{\partial z}e_{z,1}\bigg]} {8},\quad
e_{\theta,4 }=- \frac{ \bigg[5 k b_{z,3}+M\frac{\partial}{\partial z}e_{z,3}\bigg]} {24},\nonumber \\
e_{r,2}&= -\frac{ \bigg[ M k b_{z,1}+3 \frac{\partial}{\partial z}e_{z,1}\bigg]} {8  },\quad
e_{r,4}= -\frac{ \bigg[ M k b_{z,3}+5 \frac{\partial}{\partial z}e_{z,3}\bigg]} {24  }.\nonumber \\
 b_{\theta,0}&=-M b_{r,0},\nonumber\\
 b_{\theta,2 }&= \frac{ \bigg[-3 k e_{z,1}+M\frac{\partial}{\partial z}b_{z,1}\bigg]} {8},\quad
 b_{\theta,4 }=\frac{ \bigg[-5 k e_{z,3}+M\frac{\partial}{\partial z}b_{z,3}\bigg]} {24},\nonumber \\
 b_{r,2}&= -\frac{ \bigg[ -M k e_{z,1}+3 \frac{\partial}{\partial z}b_{z,1}\bigg]} {8  },\quad
 b_{r,4}= -\frac{ \bigg[ -M k e_{z,3}+5 \frac{\partial}{\partial z}b_{z,3}\bigg]} {24}.\nonumber 
 \end{align}

\begin{figure}[htbp]
	\centering
	\includegraphics*[height=55mm]{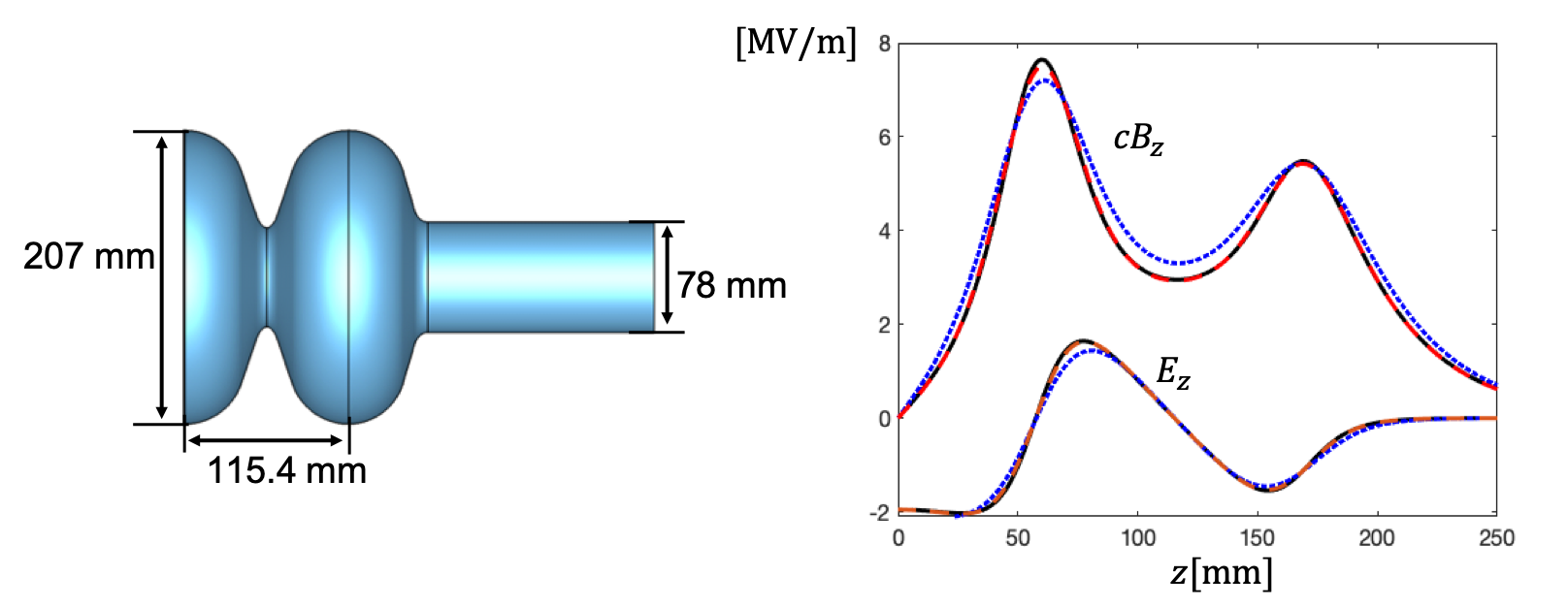}
	\caption{The left plot sketches the RF gun cavity. The right plot  shows the longitudinal field components at the transverse position $x=15$ mm, $y=15 $ mm. The blue dotted curves correspond to $N=1$ coefficients. The red dashed curves  present the results for $N=2$ coefficients.}\label{Fig004}
\end{figure}

\begin{figure}[htbp]
	\centering
	\includegraphics*[height=55mm]{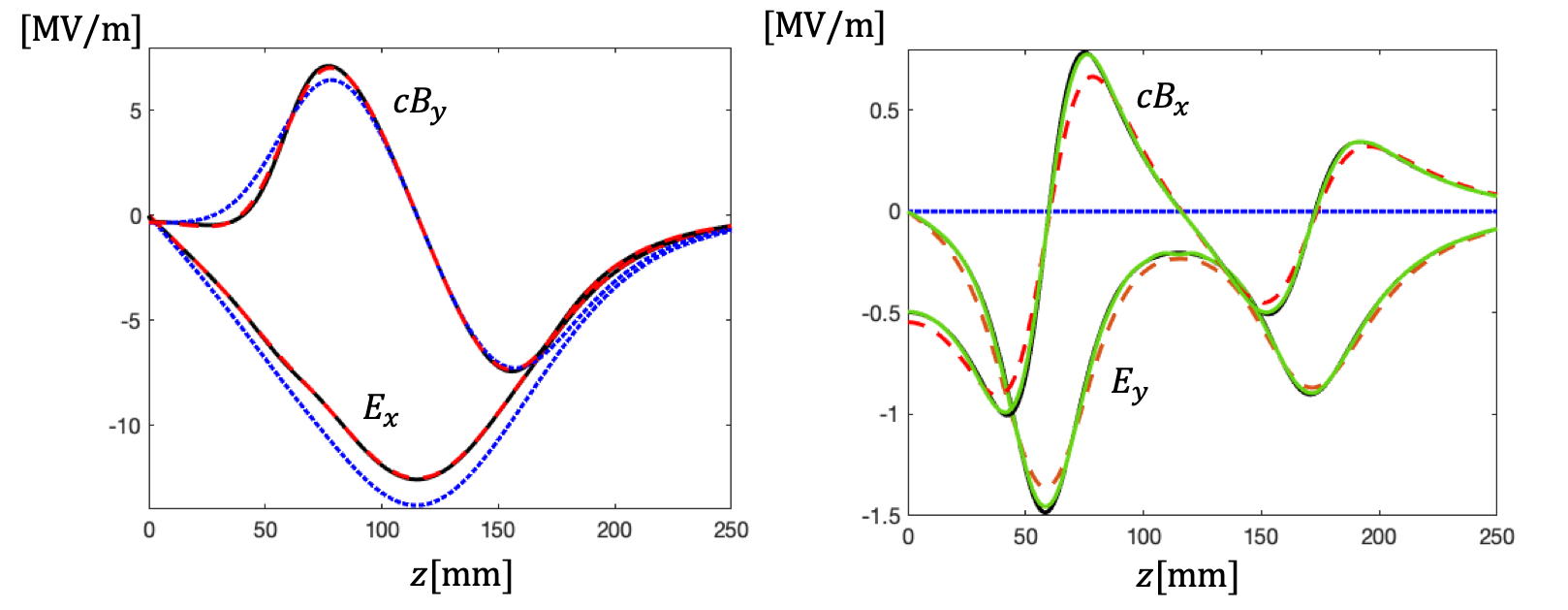}
	\caption{The plots show the transverse field components of the dipole mode in rf gun at the transverse position $x=15$ mm, $y=15$ mm. The blue dotted curves are obtained with $N=1$ coefficient. The red dashed curves  present the results for $N=2$ coefficients. The green solid curves in the right plot correspond to $N=3$ coefficients.}\label{Fig005}
\end{figure}

As a first example we consider the dipole $TM_{132}$ mode~\cite{Sym} in a perfectly conducting spherical resonator with radius $a$. The radial field components on the axis are:
\begin{align} 
E_r(0,z)=-\frac{6}{z}\frac{\partial}{\partial z}\bigg(\sqrt{k z}J_{3\frac{1}{2}}(kz) \bigg),\quad 
B_r(0,z)=-\frac{6 k^2}{\sqrt{k z}}J_{3\frac{1}{2}}(k z),
\quad k=\frac{8.7218}{a}.\nonumber
\nonumber 
\end{align}
The azimuathal magnetic  field component $B_{\theta}^{(6)}$ calculated by Eq.~(\ref{DipTransB}) with  $N=6$ terms is presented in the left plot of Fig.~\ref{Fig003}. The relative  error of the approximation
\begin{align} \label{de}
\delta&=\frac{|B_{\theta}-B_{\theta}^{(6)}|}{max(|B_{\theta}|)},\\
B_{\theta}(\rho,\psi)&=-\frac{3 k}{8\rho}\sqrt{k \rho} J_{3\frac{1}{2}}(k \rho)\bigg(\cos(\psi)+15 \cos(3 \psi)\bigg),\nonumber\\
\rho&=\sqrt{r^2+z^2},\quad \psi=\arctan(r/z)+0.5 \pi (1-z/|z|),
\nonumber 
\end{align}
is shown in the right plot in Fig.~\ref{Fig003} up to radius $r=0.5a$.

As the last example we consider a dipole mode ("$M^-$") in the rf gun cavity sketched in Fig.~\ref{Fig004}. The dipole mode of frequency  $f=1.68483$ GHz was calculated numerically. The fields are scaled such that the total energy in the entire computational domain is 1 Joule. We consider the field components, Eq.~(\ref{EqHF}), at time $t=0.125/f$ and at the transverse position $x=15$ mm, $y=15$ mm.

The right plot in Fig.~\ref{Fig004} shows the longitudinal fields. The black solid curves present the fields from the electromagnetic solver~\cite{Ack}. The dotted blue curves are the approximations obtained with only one term, $N=1$, in Eq.~(\ref{DipLong}). The dashed red curves are the approximations with two terms, $N=2$. They are already very close to the solid black curves.

The plots in Fig.~\ref{Fig005} show the transverse field components. The black solid curves present the field components from the electromagnetic solver. The dotted blue curves are the approximations obtained with only one term, $N=1$, in Eq.~(\ref{DipTransE}), (\ref{DipTransB}). The dashed red curves are the approximations with two terms, $N=2$.  They are very close to the solid black curves in the left plot for field components $E_x, B_y$. However, for the components  $E_y, B_x$ in the right plot only the approximation with three coefficients, $N=3$, reproduces  the field components with the resolution of the graph. 

\section{Summary}
In this paper we have derived a representation of arbitrary electromagnetic fields near some  axis through the coaxial filed components on the axis. The obtained representation is more accurate and requires less computational efforts than the Fourier-Bessel series approach. It was confirmed by several examples that the power series representation reproduces the electromagnetic fields accurately near to the reference orbit.

\section*{Acknowledgments}
The authors thank Wolfgang Ackermann from Technische Universität Darmstatd for calculation of the three dimensional sampling data of the dipole mode shown in Fig.~\ref{Fig004} and Fig.~\ref{Fig005}. 

\section*{Appendix  A: Representation of coefficients through initial field coefficients}

In the main text of the paper we have presented the expansion coefficients of the power series through the recursive relations. Such kind of representation 
is well suited for the computer implementation. Here we will give a non-recursive  representations, which can be used  to estimate the power series truncation error.

For m even the coefficients have the form
\begin{align} 
e_{z,2n}&= \frac{\Box^{n-m/2} e_{z,m}}{F(m,n)},\quad
F(m,n)= \prod_{p=m/2+1}^n [(2p)^2-m^2].
\nonumber 
\end{align}
and the same form of the coefficients for $b_z$ component.
The transverse field coefficients can be written as 
\begin{align}
e_{r,2n-1}= -\frac{\Box^{n-m/2-1} \bigg[M \frac{\partial}{c\partial t}b_{z,m}+2n \frac{\partial}{\partial z}e_{z,m}\bigg]} {F(m,n) },\quad 
e_{\theta,2n-1}=- \frac{\Box^{n-m/2-1} \bigg[2n\frac{\partial}{c\partial t}b_{z,m}+M\frac{\partial}{\partial z}e_{z,m}\bigg]} {F(m,n)},\nonumber \\
b_{r,2n-1}= -\frac{\Box^{n-m/2-1} \bigg[M \frac{\partial}{c\partial t}e_{z,m}+2n \frac{\partial}{\partial z}b_{z,m}\bigg]} {F(m,n) },\quad
b_{\theta,2n-1}= \frac{\Box^{n-m/2-1} \bigg[2n\frac{\partial}{c\partial t}e_{z,m}+M\frac{\partial}{\partial z}b_{z,m}\bigg]} {F(m,n) }.\nonumber 
\end{align}

Let us now rewrite the formulas for $m$ non-even. The electric field coefficients have the form
\begin{align} 
e_{z,2n+1}&= \frac{\Box^{n-(m-1)/2} e_{z,m}}{G(m,n)},\quad
G(m,n)= \prod_{p=(m-1)/2+1}^n [(2p+1)^2-m^2].\nonumber 
\end{align}
and the same form of the coefficients for $b_z$ component. The transverse field coefficients can be written as
\begin{align}
e_{r,2n}= -\frac{\Box^{n-(m-1)/2-1} \bigg[M \frac{\partial}{c\partial t}b_{z,m}+(2n+1) \frac{\partial}{\partial z}e_{z,m}\bigg]} {G(m,n) },\nonumber \\
e_{\theta,2n }=- \frac{\Box^{n-(m-1)/2-1} \bigg[(2n+1)\frac{\partial}{c\partial t}b_{z,m}+M\frac{\partial}{\partial z}e_{z,m}\bigg]} {G(m,n)},\nonumber \\
b_{r,2n}= -\frac{\Box^{n-(m-1)/2-1} \bigg[M \frac{\partial}{c\partial t}e_{z,m}+(2n+1) \frac{\partial}{\partial z}b_{z,m}\bigg]} {G(m,n)  },\nonumber \\
b_{\theta,2n}= \frac{\Box^{n-(m-1)/2-1} \bigg[(2n+1)\frac{\partial}{c\partial t}e_{z,m}+M\frac{\partial}{\partial z}b_{z,m}\bigg]} {G(m,n) }.\nonumber 
\end{align}

 \section*{Appendix  B: Analytical form of the finite length solenoid field}
 
 The axial and radial fields at any point inside of a finite
solenoid of length $L$ and radius $a$ with infinitely thin walls can be written in terms of complete elliptic integrals~\cite{Callaghan,Haas}:
\begin{align} \label{ABrSol}
B_r&= \frac{ B_z^0}{2}\frac{1}{\pi}\sqrt{\frac{a}{r}}\bigg(F(z+0.5L)-F(z-0.5L)\bigg), \tag{B1}\\
 \label{ABzSol}
B_z&= \frac{ B_z^0}{2}\frac{1}{2\pi\sqrt{a r}}\bigg(G(z+0.5L)-G(z-0.5L)\bigg),\tag{B2}\\
F(z)&=\bigg( \frac{k(z)-2}{\sqrt{k(z)}}K\big(k(z)\big)+\frac{2}{\sqrt{k(z)}}E \big(k(z) \big)\bigg),\nonumber\\
G(z)&=z \sqrt{k(z)} \bigg( K\big(k(z)\big)+\frac{a-r}{a+r}\Pi\big(h,k(z) \big)\bigg),\nonumber\\
k(z)&=\frac{4 a r}{(a+r)^2+z^2},\quad h=\frac{ 4 a r}{(a+r)^2}, \nonumber
\end{align}
where $a$ is the solenoid radius, $ B_z^0$ is the constant field of infinitely long solenoid, $K$, $E$ and $\Pi$ are closed elliptical integrals of the first, the second and the third kind, correspondingly~\cite{Abr}.

\end{document}